\documentclass{emulateapj}

\newcommand{\Msun}{M_{\odot}}
\newcommand{\Mpc}{\mathrm{Mpc}}
\newcommand{\taueff}{\tau_{\mathrm{eff}}}
\newcommand{\taues}{\tau_{\mathrm{es}}}
\newcommand{\Treion}{T_{\mathrm{reion}}}
\newcommand{\Lc}{l_\mathrm{c}}
\newcommand{\zreion}{z_{\mathrm{reion}}^{\mathrm{inst}}}

\newcommand{\MFP}{\lambda_{\mathrm{mfp}}^{912}}

\newcommand{\zstart}{z_{\mathrm{start}}}
\newcommand{\zend}{z_{\mathrm{end}}}

\newcommand{\Mmin}{M_{\mathrm{min}}}
\newcommand{\CPDF}{P(<\taueff)}

\newcommand{\HI}{H{\sc~i}}
\newcommand{\HII}{H{\sc~ii}}
\newcommand{\HeI}{He{\sc~i}}

\shorttitle{Large Opacity Variations in the High-$z$ Ly$\alpha$ Forest}
\shortauthors{D'Aloisio et al.}

\begin{document}

\title{Large opacity variations in the high-redshift Ly$\alpha$ forest: \\ the signature of relic temperature fluctuations from patchy reionization}

\author{Anson D'Aloisio\altaffilmark{1}$^\dagger$, Matthew McQuinn\altaffilmark{1}, \& Hy Trac\altaffilmark{2}}
\email{$\dagger$ anson@u.washington.edu}

\altaffiltext{1}{Astronomy Department, University of Washington, Seattle, WA 98195}
\altaffiltext{2}{McWilliams Center for Cosmology, Department of Physics, Carnegie Mellon University, Pittsburgh, PA 15213}

\begin{abstract}
Recent observations of the Ly$\alpha$ forest show large-scale spatial variations in the intergalactic Ly$\alpha$ opacity that grow rapidly with redshift at $z>5$, far in excess of expectations from empirically motivated models.  Previous studies have attempted to explain this excess with spatial fluctuations in the ionizing background, but found that this required either extremely rare sources or problematically low values for the mean free path of ionizing photons.  Here we report that much -- or potentially all -- of the observed excess likely arises from residual spatial variations in temperature that are an inevitable byproduct of a patchy and extended reionization process.    The amplitude of opacity fluctuations generated in this way depends on the timing and duration of reionization.  If the entire excess is due to temperature variations alone, the observed fluctuation amplitude favors a late-ending but extended reionization process that was roughly half complete by $z\sim9$ and that ended at $z\sim6$.  In this scenario, the highest opacities occur in regions that reionized earliest, since they have had the most time to cool, while the lowest opacities occur in the warmer regions that reionized most recently.  This correspondence potentially opens a new observational window into patchy reionization.
\end{abstract}


\section{Introduction}

When the first galaxies emerged $\approx100 - 500$ million years after the Big Bang, their starlight reionized and heated the intergalactic hydrogen that had existed since cosmological recombination.    Much is currently unknown about this process, including what spatial structure it had, when it started and completed, and even which sources drove it.  The Ly$\alpha$ forest provides one of the only robust constraints on this process, showing that it was at least largely complete by $z\approx6$, when the Universe was one billion years old \citep{fan06, 2008MNRAS.386..359G, mcgreer15}.  This Letter argues that there exists another, potentially groundbreaking signature of reionization in the Ly$\alpha$ forest data.  

The amount of absorption in the Ly$\alpha$ forest can be quantified by the effective optical depth, $\taueff\equiv-\ln\langle F\rangle_L$, where $F\equiv\exp[-\tau_{\rm Ly\alpha}(x)]$ is the transmitted fraction of a quasar's flux, $\langle...\rangle_L$ indicates an average over a segment of the forest of length $L$, and $\tau_{\rm Ly\alpha}(x)$ is the optical depth in Ly$\alpha$ at location $x$ along a sightline.  The optical depth, $\tau_{\rm Ly\alpha}(x)$, scales approximately as the \HI\ number density, which after reionization scales as $T^{-0.7}\Delta_b^2/\Gamma$.  Here, $T$ is temperature, $\Delta_b(x)$ is the gas density in units of the cosmic mean, and $\Gamma(x)$ is the \HI\ photoionization rate, which scales with the amplitude of the local ionizing radiation background.  

Observations of high-$z$ quasars show a steep increase in the dispersion of $\taueff$ among coeval forest segments around $z=6$  \citep{fan06,2015MNRAS.447.3402B}.  In the limit of a uniform ionizing background, the well-understood fluctuations in $\Delta_b$ fall well short of producing the observed dispersion at $z\gtrsim5.5$, as shown recently by \citet{2015MNRAS.447.3402B} (hereafter B2015).  Previous studies have attempted to explain this excess with spatial fluctuations in the ionizing background.  The properties of spatial fluctuations in the background depend on the number density of sources and the mean free path of photons, $\MFP$.  While $\MFP$ is well constrained at $z<5.2$, being too large to yield significant background fluctuations for standard source models \citep{2014MNRAS.445.1745W}, B2015 showed that the excess $\taueff$ dispersion at $z=5.6$ could be matched in a model where $\MFP$ decreases by a factor of $\approx 5$ between $z=5.2$ and $z=5.6$ -- a time scale of just $100$ million years.  However, such rapid evolution in $\MFP$ is inconsistent with extrapolations based on measurements at lower redshifts \citep{becker13,2014MNRAS.445.1745W}, and would imply that the emissivity of ionizing sources, in turn, increases by an unnatural factor of $\approx 5$ over the same cosmologically short time interval.  Because of these issues, B2015 speculated that the excess dispersion was evidence for large spatial variations in the mean free path\footnote{See also \citet{2015arXiv150907131D}, which appeared after we submitted this paper.}.  Alternatively, fluctuations in the ionizing background could have been enhanced if the sources of ionizing photons were rarer than the observed population of galaxies.  However, current models require half of the background to arise from bright sources with an extremely low space density of $\sim10^{-6}~\Mpc^{-3}$\citep{2015arXiv150501853C}.  This scenario is a possibility of current debate \citep[e.g.][but see D'Aloisio et al. in prep.]{2015arXiv150707678M}.  

In this Letter, we explore a source of dispersion in $\taueff$ that has so far been neglected and that, unlike ionizing background fluctuations, has straightforward implications for the reionization process itself.  In addition to $\Delta_b(x)$ and $\Gamma(x)$, the Ly$\alpha$ opacity depends on $T(x)$, mainly because the amount of neutral hydrogen after reionization is proportional to the recombination rate, which scales as $T^{-0.7}$.  Previous attempts to model Ly$\alpha$ opacity fluctuations had not included the residual temperature fluctuations that must have been present if reionization were patchy and temporally extended.  As ionization fronts propagated supersonically through the IGM, the gas behind them was heated to tens of thousands of degrees Kelvin by photoionizations of \HI\ and \HeI.  After reionization, the gas cooled mainly through adiabatic expansion and through inverse Compton scattering with cosmic microwave background (CMB) photons \citep{1994MNRAS.266..343M,1997MNRAS.292...27H,2003ApJ...596....9H,2015arXiv150507875M}.  Since different regions in the IGM were reionized at different times, these heating and cooling processes imprinted an inhomogeneous distribution of intergalactic temperatures that persisted after reionization \citep{2008ApJ...689L..81T,2009ApJ...706L.164C,furlanetto09,2014ApJ...788..175L}.  We will show that these residual temperature variations likely account for much of the observed dispersion in $\taueff$ at $z\gtrsim 5.5$, and may even account for all of it -- a scenario that would yield new information on the timing, duration, and patchiness of reionization. 

The remainder of this Letter is organized as follows.   In \S \ref{SEC:methods} and \S \ref{SEC:toymodel}, we describe our simulations and methodology.  In \S \ref{SEC:results}, we present our main results.  In \S \ref{SEC:conclusion}, we offer concluding remarks.  We use comoving units for distances and physical units for number densities.

\section{Numerical Methods}
\label{SEC:methods}

\begin{figure*}
\includegraphics[width=8.5cm,clip]{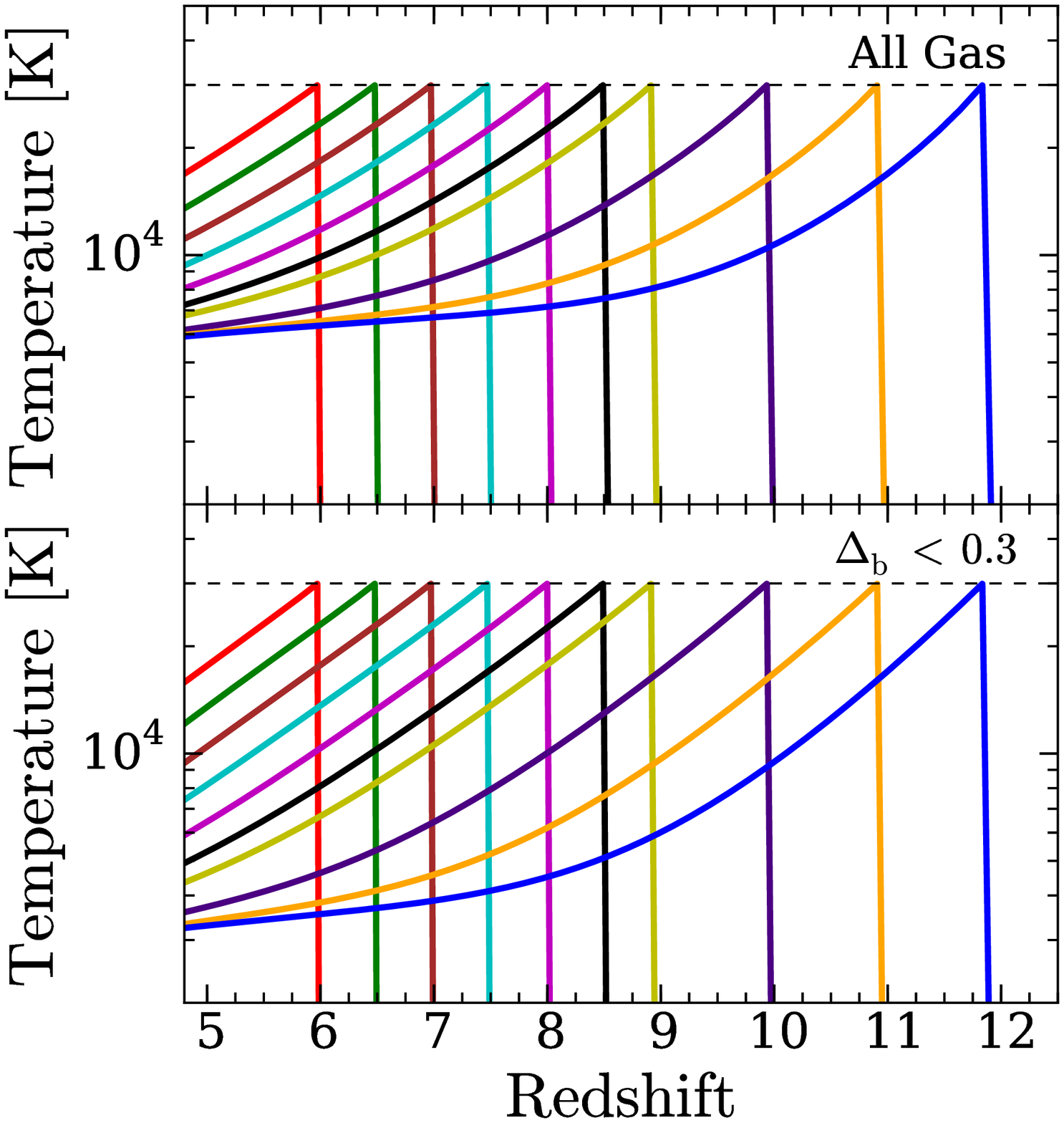}
\includegraphics[width=8.5cm,clip]{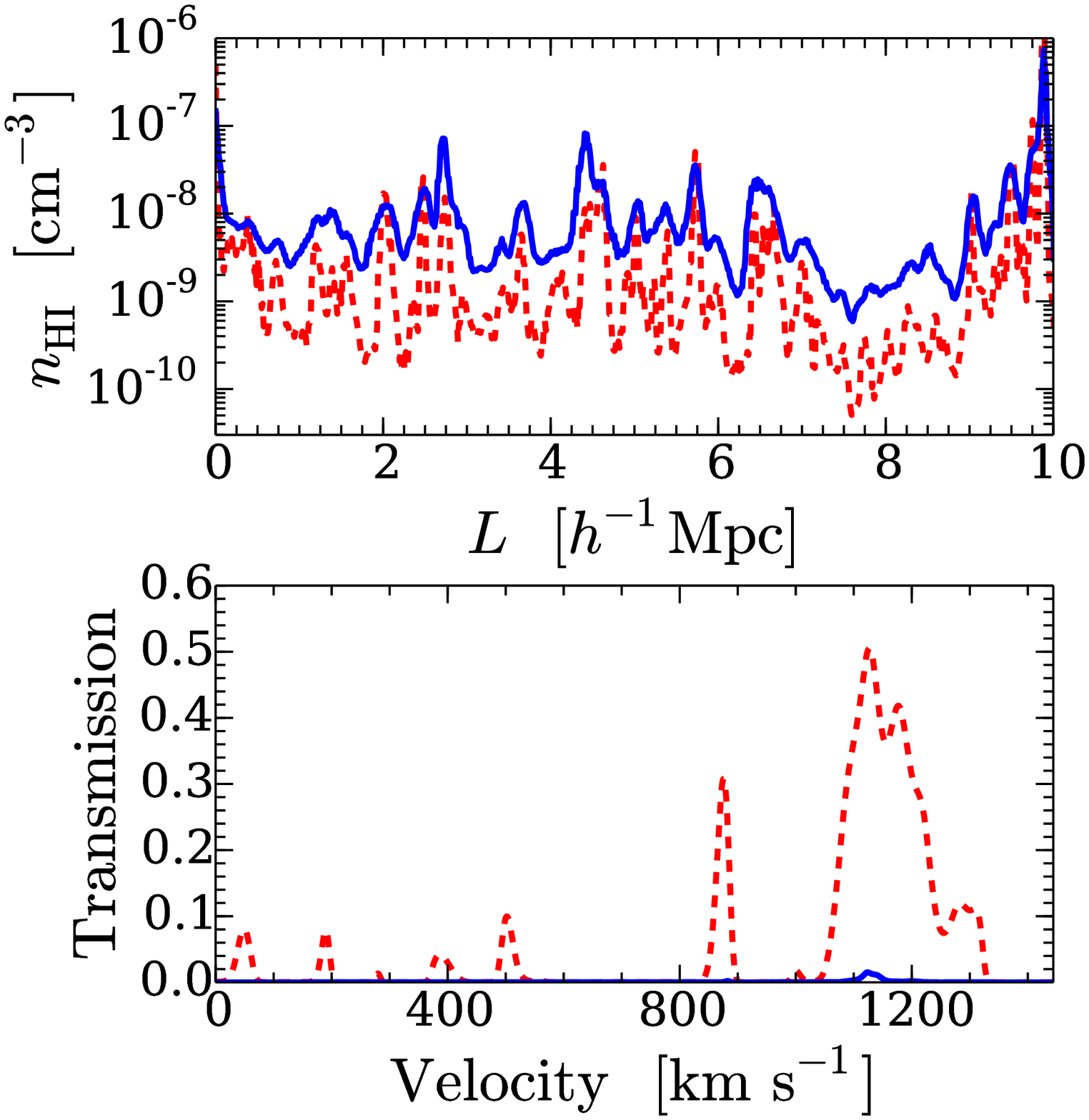}
\caption{ Illustration of the thermal and hydrodynamic relaxation of intergalactic gas after reionization and of how this relaxation affects the Ly$\alpha$ forest opacity. {\bf Left panels:}  Evolution of the volume-weighted average gas temperature in our set of ten hydrodynamical simulations in which reionization is instantaneous at $\zreion$. Different curves correspond to different $\zreion$.  In each simulation, the IGM is heated to a temperature of $\Treion=30,000~\mathrm{K}$ at redshift $\zreion$.  {\bf Right panels:}  Two example $10h^{-1}~\Mpc$ skewer segments at $z=5.8$ from the simulations with $\zreion=12$ (blue/solid curve) and $\zreion=6$ (red/dashed).  The two segments are drawn from the same location in the simulation box so they have nearly identical underlying dark matter fields.  The top panel shows the physical \HI\ number density, while the bottom panel shows the fraction of transmitted flux in the forest.}  
\label{FIG:physics}
\end{figure*}

To model the impact of relic temperature fluctuations from reionization on the distribution of $\taueff$, we ran a suite of 20 cosmological hydrodynamics simulations using a modified version of the code of \cite{2004NewA....9..443T}.  The simulations were initialized at $z=300$ from a common cosmological initial density field. We used a matter power spectrum generated by CAMB  \citep{Lewis:1999bs} assuming a flat $\Lambda$CDM model with $\Omega_m=0.3051$, $\Omega_b=0.04823$, $h=0.68$, $\sigma_8=0.8203$, $n_s=0.9667$, and $Y_{\mathrm{He}}=0.2453$, consistent with recent measurements \citep{2015arXiv150201589P}.  Our production runs use a cubical box with side length $L_{\mathrm{box}}=12.5h^{-1}~\Mpc$, with $N_{\mathrm{dm}}=1024^3$ dark matter particles and $N_{\mathrm{gas}}=1024^3$ gas cells.  

In each simulation, reionization was modeled in a simplistic manner by instantaneously ionizing and heating the gas to a temperature $\Treion$ at a redshift of $\zreion$.  Subsequently, ionization was maintained with a homogeneous background with spectral index $\alpha=0.5$, consistent with recent post-reionization background models \citep{2012ApJ...746..125H}.  Utilizing the periodic boundary conditions of our simulations, we trace skewers of length $L=50h^{-1}~\Mpc$ (following the convention of the B2015 $\taueff$ measurements) at random angles through all of the hydro simulation snapshots.  Each skewer is divided into $N_x=4096$ equally spaced velocity bins of size $\Delta v_\mathrm{skewer} = \dot{a} L/N_x$ (where $a$ is the cosmological scale factor), and Ly$\alpha$ optical depths are computed using the method of \cite{1998MNRAS.301..478T}.  Although reionization occurs instantaneously at $\zreion$ within each simulation box, we piece together skewer segments from simulations with different $\zreion$ to model the effect of an inhomogeneous reionization process, as we describe further in the next section.   

The post-reionization temperatures in the simulations are relatively insensitive to the spectrum of the ionizing background, but they are sensitive to the amount of heating that is assumed to occur at the time a gas parcel is reionized.  Previous calculations \citep{1994MNRAS.266..343M,2008ApJ...689L..81T,mcquinn-Xray} have bracketed the range of possible reionization temperatures to $\Treion \approx 20,000-30,000$ K.  (We note that previous large-scale reionization simulations do not accurately capture $\Treion$, as they do not resolve the $\sim0.3~$ physical kpc ionization fronts.)  Thus, we have run two sets of ten simulations -- one set with $\Treion=20,000$ K and the other with $\Treion=30,000$ K -- where each set contains instantaneous reionization redshifts of $\zreion=\{6,6.5,7,7.5,8,8.5,9,10,11,12\}$.  This redshift range spans the likely duration of reionization.  Simulations with $\zreion\gtrsim12$ are driven to a common temperature by $z<6$, so they are well approximated by the $\zreion=12$ simulation.   

Figure \ref{FIG:physics} shows the post-reionization thermal and associated hydrodynamic relaxation of intergalactic gas (and its effect on the Ly$\alpha$ forest opacity) in our simulations.  The top-left panel shows the volume-weighted average gas temperature in the ten $\Treion=30,000$ K simulations.  The bottom-left panel shows this same average, but limited to gas cells with densities of $\Delta_b<0.3$, the deepest voids that dominate transmission in the highly-saturated $z\gtrsim5$ forest.  Even at $z\sim5$, the $\Delta_b <0.3$ gas temperatures differ by up to a factor of five between the simulations that were reionized at different times.  The right panels show the \HI\ number densities (top) and the transmission (bottom) at $z=5.8$ for the same skewer through our $\zreion=12$ (blue/solid) and $\zreion=6$ (red/dashed) simulations.  The transmission is nearly zero in the $\zreion=12$ case, owing to colder temperature and hence enhanced \HI\ densities, whereas there is significant transmission in the $\zreion=6$ case.

\section{Constructing the $\taueff$ distribution: a toy model}
\label{SEC:toymodel}

Reionization is a process that is inhomogeneous and temporally extended, unlike in our individual hydro simulations.  Modeling the thermal imprint of patchy reionization on the Ly$\alpha$ forest thus requires an additional ingredient: a model for the redshifts at which points along our skewers are reionized.  In this section, we present a simplified toy model to illustrate how we piece together skewer segments from our hydro simulations, and to provide insight into how the timing, duration and morphology of reionization affect the amplitude of Ly$\alpha$ opacity fluctuations.  
   
For illustrative purposes, let us assume that the Ly$\alpha$ forest is made up of segments of equal length, $l_c$, where each segment is reionized at a single redshift.  Let us further assume that the reionization redshift of each segment is drawn from a uniform probability distribution over the interval $[\zstart, \zend]$, resulting in a global reionization history in which the mean ionized fraction, $\bar{x}_{\mathrm{HII}}(z)$, is linear in redshift (since $\bar{x}_{\mathrm{HII}}(z)$ is the cumulative probability distribution of the reionization redshift). 

Assuming that the reionization redshift field is in the Hubble flow, each segment of length $l_c$ spans $\Delta N_x=4096\times l_c/(50h^{-1}\Mpc)$ velocity bins of our hydro simulation skewers.  For the first segment, with reionization redshift $z_1$, we take the initial $\Delta N_x$ spectrum velocity bins of a randomly drawn skewer that has $\zreion$ closest to $z_1$.  For the next segment, with reionization redshift $z_2$, we take the next $\Delta N_x$ velocity bins \emph{from the same skewer} through the simulation that has $\zreion$ closest to $z_2$ (note that all simulations were started from the same initial density field).  We repeat this procedure until an entire $50h^{-1}~\Mpc$ sightline is filled. 

In what follows, we compare these toy models against the B2015 measurements of the cumulative probability distribution function of $\taueff$, $\CPDF$.  The measurements considered here span $z=5.1-5.9$ in bins of width $\Delta z=0.2$.  We construct $\CPDF$ from 4000 randomly drawn sightlines at the central redshift of each bin.    For each redshift, we rescale the nominal post-reionization photoionization rate of our simulations ($\Gamma=10^{-13}~\mathrm{s}^{-1}$) by a constant factor, such that our model $\CPDF$ is equal to the observed distribution at either $\CPDF=0.15$ or $\CPDF=0.3$, depending on which value provides the better visual fit.  We have performed extensive numerical convergence tests using simulations of varying resolution and box size.  We found excellent convergence of $\CPDF$ for our production runs in both box size and resolution (especially for our patchy reionization models).     

 \begin{figure}
\begin{center}
  \hspace{-0.5cm}
\includegraphics[width=9cm,clip]{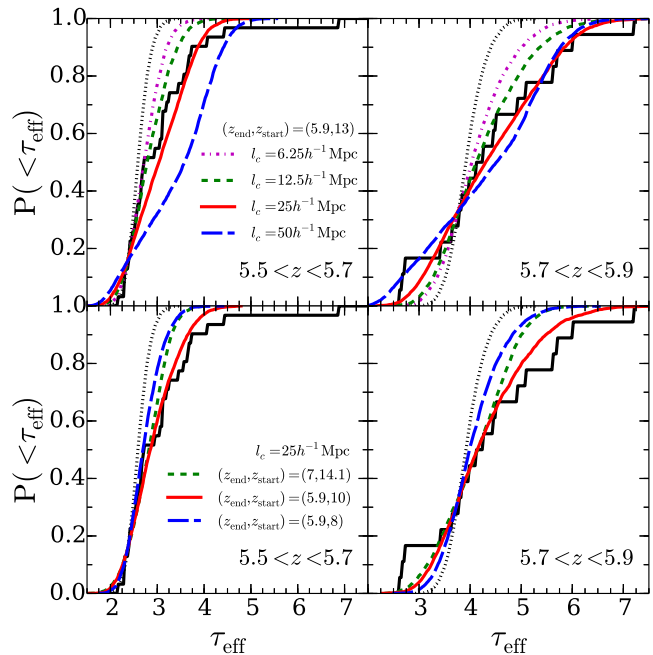}
\end{center}
\caption{Comparison of cumulative probability distribution functions of $\taueff$ in our toy model (colored curves) compared to the B2015 measurements (black histograms) and to an instantaneous reionization model with $\zreion=8.5$ (black/dotted curves).  {\bf Top row:}  Varying the coherence length, $\Lc$, assuming a uniform $\zreion$ probability between $z=5.9-13$. {\bf Bottom row:} Varying the redshift interval over which the $\zreion$ values are drawn, assuming $\Lc=25h^{-1}~\Mpc$. }
\label{FIG:toymodel_CDFs}
\end{figure}
        
Figure \ref{FIG:toymodel_CDFs} shows the $\CPDF$ of these toy models for a range of $(\zstart,\zend,l_c)$, compared against the B2015 measurements in the two highest redshift bins (black histograms).   The curves with shorter reionization durations, or with smaller $l_c$, fall closer to the black/dotted curves, which assume that reionization occurs instantaneously (at $\zreion=8.5$, although it does not depend on this choice).  Figure \ref{FIG:toymodel_CDFs} affords three insights: (1) The larger the coherence length, $l_c$, over which gas shares a similar reionization redshift, the larger the spread in $\taueff$; (2) The width of the observed $\CPDF$ can be fully accounted for only if reionization ended at $z\lesssim7$ and was well underway by $z\sim9$;   (3) The maximal width is achieved by a late-ending ($\zend\sim6$) and extended reionization model in which large contiguous segments ($\gtrsim12.5h^{-1}~\Mpc$) of the Ly$\alpha$ forest were reionized at $z\gtrsim9$.  In the next section, we apply these insights to construct more realistic models of reionization that reproduce the observed width of $\CPDF$. 

\section{The Effect of Temperature Fluctuations on the High-Redshift Ly$\alpha$ Forest}  
\label{SEC:results}

\subsection{Results}

\begin{figure*}
\centerline{
\includegraphics[width=8.2cm,clip]{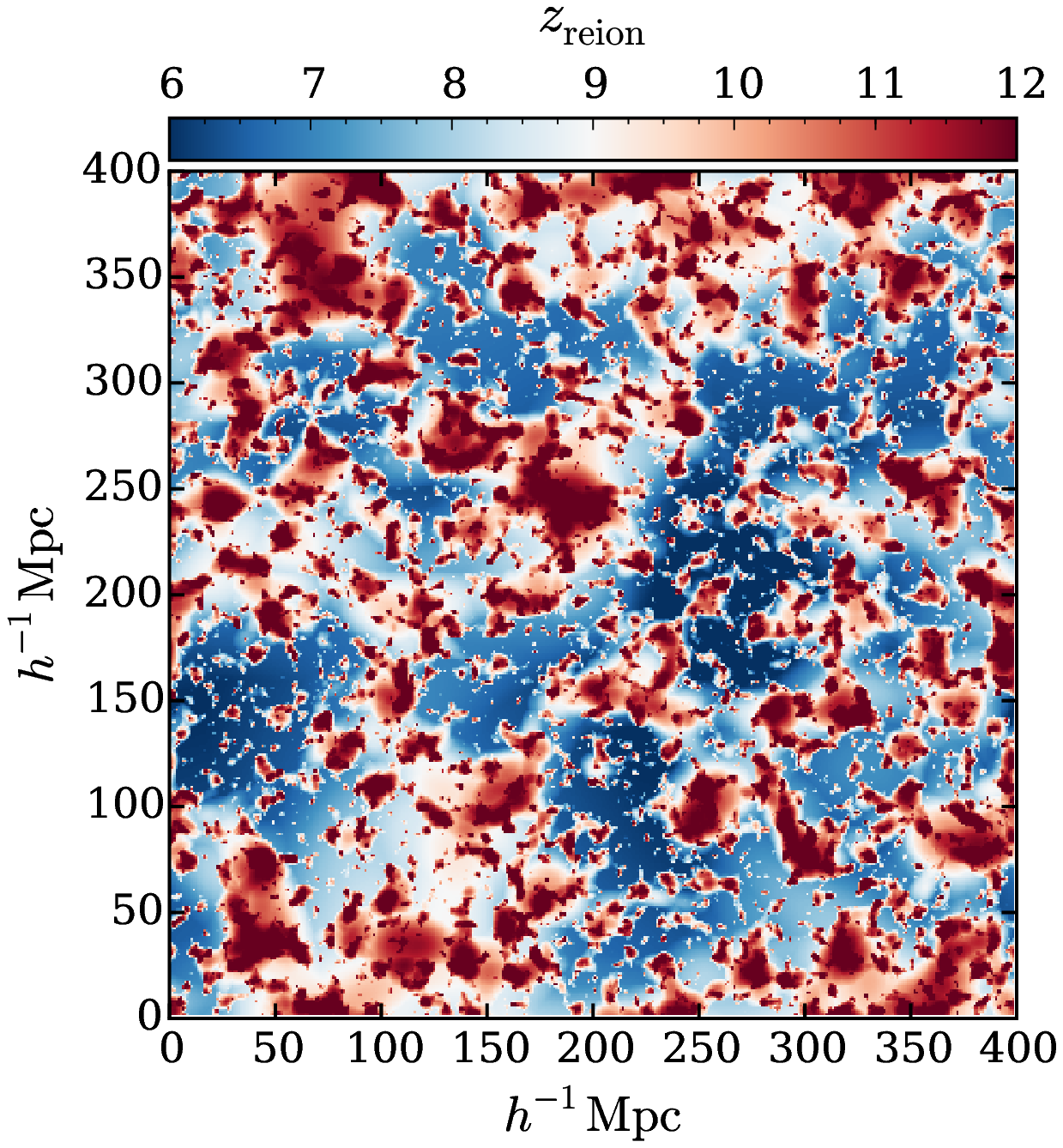}
\includegraphics[width=8.2cm,clip]{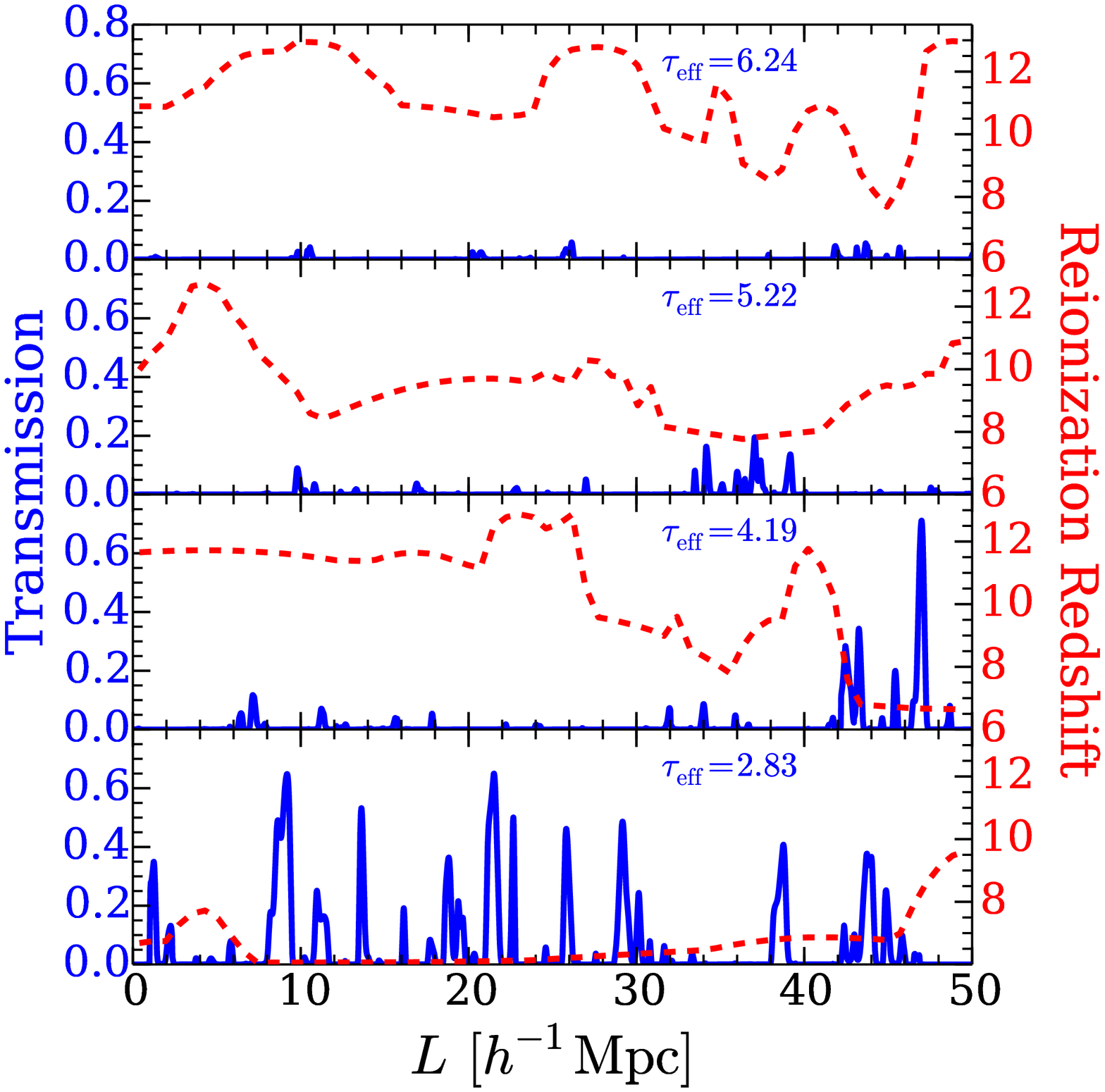}
}
\caption{{\bf Left panel:} Two-dimensional slice through the reionization redshift field of our fiducial excursion-set model.  {\bf Right panels:} Each panel shows a $50h^{-1}~\Mpc$ skewer through the fiducial model, with the four skewers selected to span a range of $\taueff$ values.  The red/dashed curves show the reionization redshift values along the skewers, and the blue/solid curves show the Ly$\alpha$ forest transmission at $z=5.8$.  The transmission curves are constructed by piecing together mock spectrum segments from our suite of hydrodynamical simulations, matching $\zreion$ to the reionization redshifts from the excursion-set model.}
\label{FIG:zreionfield}
\end{figure*}

\begin{figure*}
\begin{center}
\includegraphics[width=12.2cm,clip]{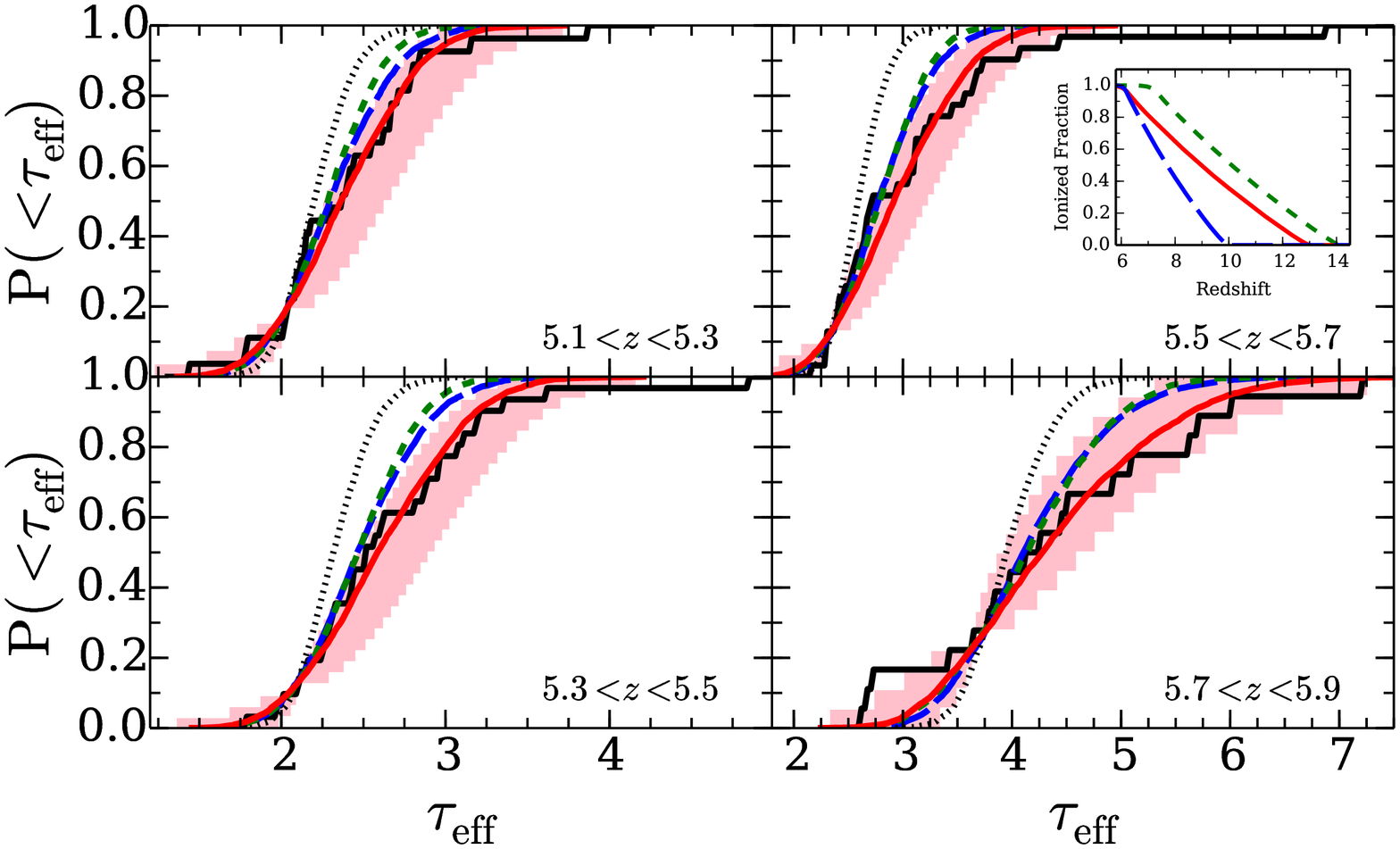}
\hspace{-0.5cm}
\includegraphics[width=6.1cm]{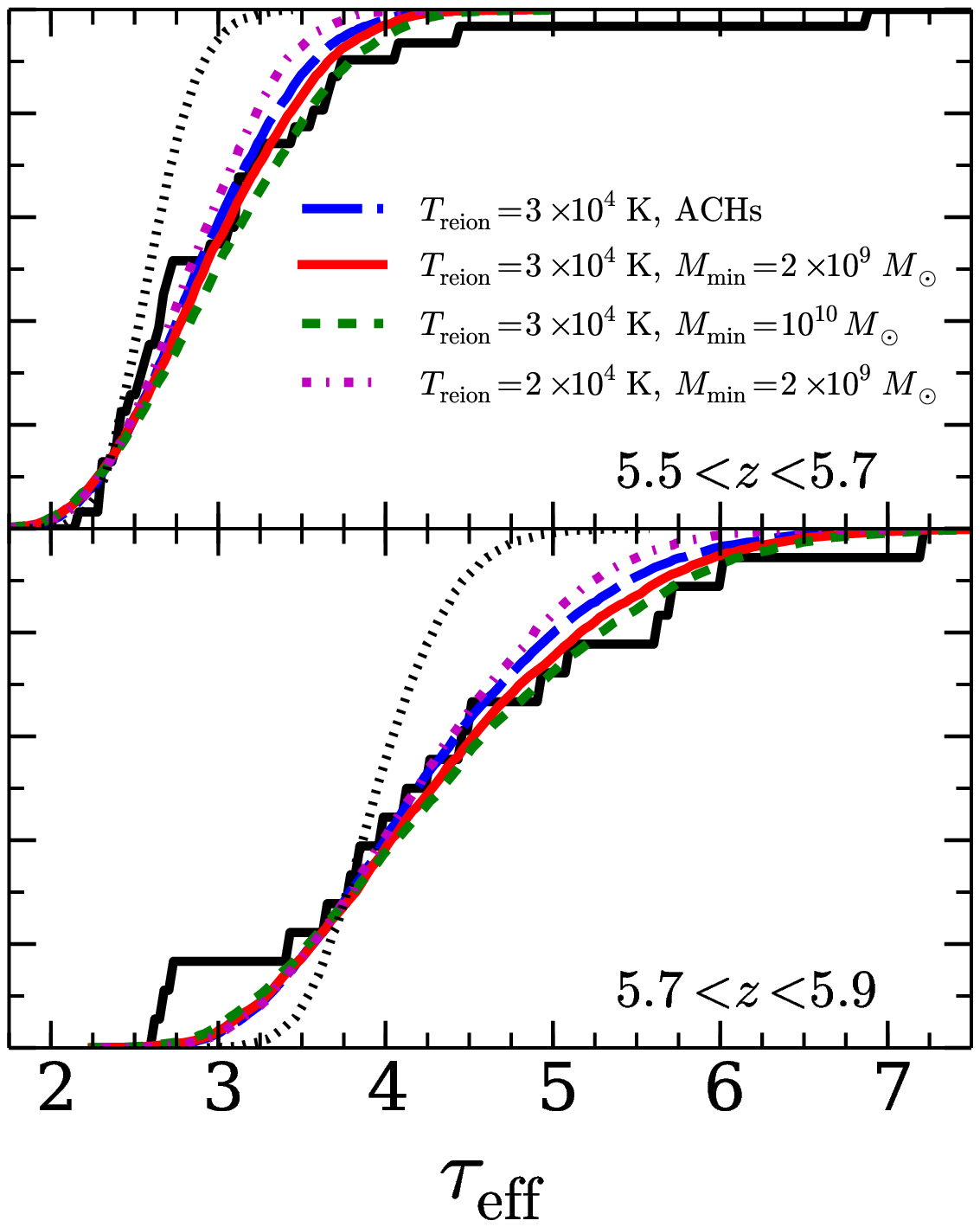}
\end{center}
\caption{{\bf Leftmost four panels:} Cumulative probability distribution functions of $\taueff$ measured in B2015 (black histograms) compared against an instantaneous reionization model (dotted curves) and against three excursion-set models of patchy reionization (other curves).  The corresponding global ionization histories of these models are shown in the sub-panel.  The pink shaded regions indicate the 90\% confidence levels of our fiducial model (red curves) in the four redshift bins shown, estimated from bootstrap realizations. {\bf Rightmost two panels:}  The effect of varying the reionization temperature, $\Treion$, and the minimum mass of halos that host galactic sources, $\Mmin$, on $\CPDF$ at the two highest redshifts reported in B2015. }
\label{FIG:ePS_CDFs}
\end{figure*}

We generate physically-motivated reionization redshift fields using simulations based on the excursion-set model of reionization (ESMR) \citep{2004ApJ...613....1F,2007ApJ...654...12Z,2009ApJ...703L.167A,2010MNRAS.406.2421S,2011MNRAS.411..955M}, which has been shown to reproduce the ionization structure found in full radiative transfer simulations \citep{2005ApJ...630..657Z, 2011MNRAS.414..727Z,2014MNRAS.443.2843M}.    In particular, our ESMR uses a realization of the linear cosmological density field and a top hat in Fourier space filter to generate a realization of reionization in a cubical box with side $400h^{-1}~\Mpc$, sampled with $512^3$ cells.  For details of the algorithm, see \citet{2007ApJ...654...12Z}.  We use the simplest formulation of the ESMR, with two free parameters: (1) The minimum mass of halos that host galactic sources of ionizing photons, $\Mmin$; (2) The ionizing efficiency of the sources, $\zeta$.   We tune $\zeta$ to obtain a reionization history that is approximately linear in redshift, similar to those in our toy model. 

The ESMR calculation yields a cube of reionization redshifts.  We construct mock absorption spectra by first tracing $50h^{-1}~\Mpc$ skewers through this cube.  We then piece together spectrum segments from our suite of hydro simulations, much like in our toy model, except here we match $\zreion$ to the reionization redshifts along the ESMR skewers.  This process does not account for how $\zreion$ correlates with density, an effect that will underestimate (making our models conservative) the width of $\CPDF$, which we address shortly.

The left panel of Figure \ref{FIG:zreionfield} shows a 2D slice through our fiducial reionization redshift field in which reionization spans $z=6-13$.   In the right panels, the blue/solid curves show the fraction of transmitted flux along four different sightlines through this field, selected to span a range of $\taueff$.  The red/dashed curves show the corresponding reionization redshifts along the sightlines.  On average, regions that reionize at later times yield more transmission, while regions that reionize at $z\gtrsim9$ result in dark gaps in the Ly$\alpha$ forest.  The variation among these mock spectra is similar to the well-known variation seen in observations of the $z\sim6$ Ly$\alpha$ forest \citep{fan06}.  In our interpretation, this variation reflects differences in the reionization redshifts -- and hence temperatures\footnote{We find that pressure smoothing plays an insignificant role in generating $\taueff$ fluctuations.} -- between segments of the Ly$\alpha$ forest.  

The leftmost four panels of Figure \ref{FIG:ePS_CDFs} show $\CPDF$ in three  ESMR models that take $\Treion=30,000~\mathrm{K}$ and $\Mmin=2\times10^9~\Msun$, with reionization histories shown in the inset of the $5.5<z<5.7$ panel.  The CMB electron scattering optical depths in these models are $\taues = 0.054, 0.068$ and $0.080$, within uncertainties of the latest {\it Planck} measurement of $\tau_{\mathrm{es}}=0.066\pm0.016$ \citep{2015arXiv150201589P}.  These models are compared against the B2015 measurements (black histograms) and against the homogeneous reionization reference model with $\zreion=8.5$ (dotted curves).  The blue/long-dashed and green/short-dashed curves correspond to scenarios in which reionization spans $z=6-10$ and $z=7-14$, respectively.  While these two models produce significantly more width in $\CPDF$ than the homogeneous reionization model, they fall short of producing the full range of $\taueff$ required to match the observations.   However, the fiducial $z=6-13$ reionization model (red/solid curves) generally provides a good match to the measurements.  A significant success of the fiducial model is that the observed redshift evolution of the $\CPDF$ width is reproduced without any additional tuning of parameters.  (We have checked that this success also holds over $z=4-5$, lower redshifts than those shown where the effect is reduced.)  Indeed, there are not any parameters that can be tuned in our model to change the post-reionization evolution of the width.  The pink shaded regions indicate the 90\% confidence levels of our fiducial model, estimated from bootstrap realizations, showing consistency with all the data aside from a single high-opacity point at $z=5.4$ and at $z=5.6$.  

The discrepancy at the highest opacities may arise because our method of constructing mock absorption spectra does not capture correlations between temperature and density that should be present, since denser regions are more likely reionized earlier.  Such correlations would act to increase the width of $\CPDF$, as the denser regions around galaxies are ionized earlier in our models.  One might naively think these correlations are small, because correlations between the density on the much larger scales of the \HII\ bubbles and the smaller scales of voids in the forest should be weak, but calculations show that they may not be negligible \citep{furlanetto09, 2008ApJ...689L..81T,2013ApJ...776...81B} 

\subsection{Effect of varying reionization model parameters}

The right panels of Figure \ref{FIG:ePS_CDFs} show the effect of varying $\Treion$ and $\Mmin$.  For $\Treion=20,000~\mathrm{K}$, the distribution of $\taueff$ is somewhat narrower than the case with $\Treion=30,000~\mathrm{K}$.  Hotter temperatures are likely achieved towards the end of reionization, when \HII\ bubbles are larger and propagate at quicker speeds.  A model with $\Treion\sim30,000~\mathrm{K}$ near the end of reionization, and smaller temperatures earlier on, would likely produce more width than a model with  $\Treion\sim30,000~\mathrm{K}$ at all times.  However, any conclusions about $\Treion$ prior to modeling the density/reionization-redshift correlations are premature.

The blue/long-dashed and green/short-dashed curves in the right panels show the effect of varying $\Mmin$.  For the former atomic cooling halos (ACHs) curve, $M_{\mathrm{min}}$ is set to the minimum mass required to achieve a halo virial temperature of $10,000~\mathrm{K}$.  For both curves, we tune $\zeta$ to match the reionization history of our fiducial model (red curve in the inset).  We find that the effects of varying $M_{\mathrm{min}}$ are minor.  

\section{Conclusion}
\label{SEC:conclusion}

We have shown that residual temperature inhomogeneities from a patchy and extended reionization process likely account for much of the opacity fluctuations in the $z\gtrsim5$ Ly$\alpha$ forest.  Inhomogeneities in the ionizing background may also contribute at a significant level, though current models in this vein have required very small mean free paths or extremely rare sources.  We showed that residual temperature fluctuations alone could account for the entire spread of observed $\taueff$.  A significant success of this interpretation is that it is able to reproduce the rapid growth of $\taueff$ fluctuations with redshift, despite having very little freedom in its post-reionization evolution. In this scenario, the observations favor a late but extended reionization process that is roughly half complete by $z\sim9$ and that ends at $z\sim6$.  

Unlike ionizing background fluctuations, which do not necessarily signal the end of reionization, temperature fluctuations directly probe the timing, duration, and patchiness of this process.  If most of the opacity variations owe to temperature, it would mean that, on average, the darkest $\gtrsim10~\Mpc$ segments of the $z\gtrsim 5$ Ly$\alpha$ forest were reionized earliest, and the brightest segments last -- a potentially powerful probe of cosmological reionization.
\\ 

\acknowledgements
The authors acknowledge support from NSF grant AST1312724.  H.T.  also acknowledges support from NASA grant ATP-NNX14AB57G.  Computations were performed with NSF XSEDE allocation TG-AST140087.  We thank Paul La Plante, Jonathan Pober, Phoebe Upton Sanderbeck, George Becker, Adam Lidz, and Fred Davies for helpful discussions.

\end{document}